\documentclass[a4paper,fleqn]{mnras}
\usepackage{float}
\usepackage{amsmath}
\usepackage{booktabs}
\usepackage{pdflscape}
\usepackage{txfonts}
\usepackage[T1]{fontenc}
\usepackage{ae,aecompl}
\usepackage{graphicx}
\usepackage{color,soul}
\newcommand{\beq}{\begin{equation}}
\newcommand{\eeq}{\end{equation}}

\newcommand{\ehs}{\color{black}}
\title[Mass transfer in a planet--star system]{Instability of mass transfer in a planet--star system}

\author[Jia $\&$ Spruit]{Shi Jia$^{1,2,3,4}$\thanks{E-mail: jiashi@ynao.ac.cn}, H.C.\ Spruit$^{2}$ \\
\\
$^{1}$ Yunnan Observatories, Chinese Academy of Sciences, Kunming, 650011,China\\
$^{2}$ Max-Planck-Institut f\"{u}r Astrophysik, Karl-Schwarzschild-Str.\ 1, D-85748 Garching, Germany\\
$^{3}$ Key Laboratory for the Structure and Evolution of Celestial Objects, Chinese Academy of Sciences, Kunming 650011, China\\
$^{4}$ University of Chinese Academy of Sciences, Beijing 100049, China\\ 
}

\begin{document}

% These dates will be filled out by the publisher
\date{Accepted xxx. Received xxx; in original form xxx}

% Enter the current year, for the copyright statements etc.
\pubyear{2016}

\pagerange{\pageref{firstpage}--\pageref{lastpage}}
\maketitle

\label{firstpage}

% Abstract of the paper
\begin{abstract}
We show that the angular momentum exchange mechanism governing the evolution of mass-transferring binary stars does not apply to Roche lobe filling planets, because most of the angular momentum of the mass-transferring stream  is absorbed by the host star. Apart from a correction for the difference in specific angular momentum of the stream and the centre of mass of the planet, the orbit does not expand much on Roche lobe overflow. We explore the conditions for dynamically unstable Roche lobe overflow as a function of planetary mass and mass and radius (age) of host star and equation of state of planet. For a Sun-like host, gas giant planets in a range of mass and entropy can undergo dynamical mass transfer. Examples of the evolution of the mass transfer process are given. Dynamic mass transfer of rocky planets depends somewhat sensitively on equation of state used. Silicate planets in the range $1 \ M_{\bigoplus} <M_{\mathrm{p}} < 10 \ M_{\bigoplus} $ typically go through a phase of dynamical mass transfer before settling to slow overflow when their mass drops to less than $1 \ M_{\bigoplus}$.
\end{abstract}

% Select between one and six entries from the list of approved keywords. Don't make up new ones.

\section{Introduction}

Though constituting only a small fraction of the total exoplanet population, exoplanets orbiting close to their host stars pose interesting challenges to theoretical models for the formation and evolution of planetary systems. Since the hosts generally rotate more slowly than the  planet orbit, tidal interaction causes the planets to lose angular momentum. Depending on the somewhat uncertain strength of tidal friction (for a review see Ogilvie 2014) the planets on the closest observed orbits, on the order of a few days, would spiral into their host within a few billion years (e.g. Raymond, Barnes \& Mandell 2008; Levrard, Winisdoerffer \& Chabrier 2009; Jackson, Barnes \& Greenberg 2009). The loss of planets by spiral-in has been invoked to explain the orbital distribution of close-in exoplanets (Jackson et al. 2009) and the dearth of close-in exoplanets around fast rotating stars (Teitler \& K\"onigl 2014) from the observations (McQuillan, Mazeh \& Aigrain 2013). 

The final fate of an angular momentum losing planet depends on its mass, mean density, and composition. It also depends sensitively on the details of the angular momentum balance during the Roche overflow process, for which different assumptions have been made in previous work. In several studies it is assumed that the planet is rapidly disrupted once it fills its Roche lobe, and the material of planet is then accreted on to host star (e.g. Jackson et al. 2009; Rappaport et al. 2013; Teitler \& K\"onigl 2014). Metzger, Giannios \& Spiegel (2012), on the other hand,  found that the mass transfer of the planet-star (hot Jupiter) system will be stable, occurring on the slow tidal evolution time-scale, if $ \bar{\rho}_{\mathrm{p}} / \bar{\rho}_\star \lesssim 1 $, where $\bar{\rho}_{\mathrm{p}}$ and $\bar{\rho}_\star$ are mean density of the planet and that of the host star, respectively. 

The possible outcomes of the spiral-in process of hot Jupiter are conveniently classified into three cases (Metzger et al. 2012). With decreasing orbital separation the planet can either reach its Roche limit and disrupt before physically entering the star, or it can spiral in `whole'. In the former case the planet loses mass either as a slow process governed by the orbital evolution under tidal interaction, or dynamically, evolving on an orbital time-scale.

Whether Roche lobe overflow takes place before reaching the stellar surface depends on ratio of the mean density of the planet $\bar{\rho}_{\mathrm{p}}$ to that of the star $\bar{\rho}_\star$. If $\bar{\rho}_{\mathrm{p}} / \bar{\rho}_\star \lesssim 5 $, the planet reaches its Roche limit outside the host star. If the planet has a higher mean density, $\bar{\rho}_{\mathrm{p}} / \bar{\rho}_\star \gtrsim 5 $, it would fill its Roche lobe only below the stellar surface. In this case a direct merger occurs between the planet and the host star (Metzger et al. 2012).

As in the case of mass transfer in binary stars (Paczy{\'n}ski 1971; Frank, King \& Raine 1992), the time-scale on which Roche lobe overflow takes place depends critically on the adiabatic mass-radius relation of the planet. If loss of mass causes the radius of the planet to decrease in size more slowly than the Roche radius, mass transfer is unstable on a dynamical time-scale (`dynamical mass transfer', Paczy{\'n}ski 1965, Paczy{\'n}ski, Zi{\'o}lkowski \& Zytkow 1969). The final disruption of the planet then takes place in a few orbits (e.g.\  Rasio 1994). In the opposite case mass loss is dynamically stable. The loss rate through inner Lagrangian point (L1) is then far slower, governed by the time-scale of orbital decay through tidal interaction with the host star. 

In both cases, the material lost through the L1 either forms a disc, or hits the stellar surface and is accreted on to host star directly (as in the case of many Algol type binaries). Which is the case depends on the mass ratio $M_{\rm p}/M_*\equiv q$ of the planet's mass $M_{\rm p}$ to the star's mass $M_*$, and on the radius $R_*$ of the host star.

\section{Angular momentum}

The mass transfer between a hot Jupiter and its host has been treated (Valsecchi, Rasio \& Steffen 2014 and others) in analogy with the mass transfer in binary systems like cataclysmic variables (CVs) and X-ray binaries (as in Rappaport, Joss \& Webbink  1982, Ritter 1988), using codes for binary evolution developed for these systems. The angular momentum exchange during mass transfer is treated as approximately `conservative', in the sense that the orbital angular momentum of the companion is conserved on mass loss. The angular momentum transferred by the stream is returned to the orbit through tidal interaction of the companion with the accretion dics. 
 
This is appropriate for such systems since the angular momentum accreted by the primary is usually negligible on account of its small size, leading to the formation of a disc extending into the gravitational potential of the companion. As we will argue below in section \ref{orbtohost}, it is not an appropriate model for mass angular momentum balance in the case of planets orbiting main sequence (MS) or larger host stars.

\subsection{Evolution of the orbit}
\label{roche-orbital}

Mass through Roche lobe overflow can result from  variety of processes, including expansion of the radius of the planet (or a companion star) by internal evolution or external heating, and shrinkage of the orbit due to angular momentum loss processes such as a magnetically driven wind (`magnetic braking') or tidal interaction with its host star.

The total angular momentum of the planet-star system can be written as
\beq
\label{ltot}
J = J_{\mathrm{orb}} + J_{\mathrm{s*}} + J_{\mathrm{sp}},
\eeq
where $J_{\mathrm{orb}}$ is the orbital angular momentum of the system, $J_{\mathrm{s*}}$ and $J_{\mathrm{sp}}$ are the spin angular momentum of host star and planet, respectively. In the following, we will ignore the spin angular momentum of planet in the angular momentum balance, since for mass ratios less than $10^{-2}$ it accounts for only a fraction of a percent of the total. Tidal exchange between the host star's spin and the orbit takes place on the tidal interaction time-scale, which is slow compared with the processes we are interested in. Contrary to the case of accretion on to compact objects, the moment of inertia of the star is so large that the star can just be treated as a sink of angular momentum. 

The orbital angular momentum of the system (the total angular momentum if star and planet can be treated as point masses) is (cf. Paczy{\'n}ski 1971)
\beq J_{\rm orb}=\Omega\, a^2 \mu,\label{jorb}\eeq
where $\Omega=(G M/a^3)^{1/2}$ is the orbital angular frequency, $M=M_*+M_{\rm p}=M_*(1+q)$ is the total mass, $q=M_{\rm p}/M_*$ is the mass ratio, $a$ is the separation between the centres of mass of star and planet, and $\mu =M_*M_{\rm p}/M$ is the reduced mass. 
With the planet's spin neglected, and denoting the time derivative by an overdot ($\dot {\, \, } $), the angular momentum balance is
\beq \dot J_{\rm orb}+ \dot{J}_{\mathrm{s*}}+\dot J_{\rm ext}=0,\eeq
where $\dot J_{\rm ext}$ stands for any external toques on the system, such as the tidal torque or a stellar wind. We will ignore such torques for the moment, since they act on time-scales longer than the dynamical  mass transfer instability that is the subject here (see sections \ref{orbtohost}, \ref{discplanet} below). Angular momentum taken up by the spin of the star during the mass transfer process must be included, however.
 
With mass conservation, ($\dot M_*=-\dot M_{\rm p}$), equation (\ref{jorb}) yields
\beq {\dot J_{\rm orb}\over J_{\rm orb}}={\dot a\over 2a}+{\dot M_{\rm p}\over M_{\rm p}}(1-q), \label{dotj1}\eeq
(where $\dot M_{\rm p}<0$). The orbital change $\dot J_{\rm orb}$ is determined by the mass transfer rate and the specific angular momentum $j_{\rm s}$ that is carried with it to the host star, or is transferred back to the orbit. Write $j_{\rm s}$ as a fraction $f$ of the specific angular momentum $j_{\rm p}$ of the planet's centre of mass, which lies at a distance  $x_{\rm p}=a/(1+q)$ from the centre of mass of the system, so that
\beq j_{\rm s}=f\,\Omega\, a^2/(1+q)^2.\label{deff}\eeq
Equation (\ref{dotj1}) can then be written as
\beq {\dot a\over a}={2\over 1+q}{\dot M_{\rm p}\over M_{\rm p}}(f+q^2-1).\label{fdot}\eeq

A popular assumption has been that all the angular momentum transferred by the stream is returned to the  orbit.  (Sometimes called `angular momentum conservation'). In terms of the above, this assumption is equivalent to setting $f=0$. This is a good approximation in the case of mass transfer in CVs and X-ray binaries. In these cases the moment of inertia of the accreter can be neglected. Instead, the mass transferred forms an accretion disc around the compact object, and the angular momentum is transferred back to the orbit by tidal interaction of the donor star with the outer parts of the disc (cf. Frank, King \& Raine 1992). As a result, the orbit increases, approximately as $a\sim M_{\rm p}^{-2}$. Since this also increases the size of the Roche lobe, it has a strongly stabilizing effect on the mass transfer.  

Loss of orbital angular momentum by transfer of mass to the host is not to be confused with what is called `nonconservative' mass transfer, since it does not involve any loss of angular momentum from the system as a whole. In this sense it can be compared with orbital evolution by tidal interaction between planet and host. During dynamical instability it takes place on the much faster orbital time-scale, however. Classical nonconservative processes such as angular momentum loss in a stellar wind or to a circumbinary disc take place on  time-scales that are more important for the secular evolution of the orbit (see Valsecchi et al. 2015). 

\subsection{Angular momentum transfer from orbit to host star}
\label{orbtohost}
The moment of inertia of a MS host is so large that it can accommodate the orbital angular momentum of a Roche lobe filling planet  with only a negligible increase in spin, compared with its maximum rotation rate. 

One might consider a stream transferring mass to the host star with the specific angular momentum of the centre of mass of the planet, and absorbed by the star as an increase in rotation rate. In equation (\ref{fdot}) this would mean $f=1$. The orbital separation would not change in this case, apart from a small effect of order $q^2$  attributable to the effect the exchange of mass has on the position of the centre of mass of the system relative to star and planet. Under this assumption Roche lobe overflow would be much more likely to be dynamically unstable. 

The  specific angular momentum of the centre of mass of the planet, $j_{\rm p}=\Omega a^2/(1+q)^2$, is not the same as that of the mass lost by the planet, however, since it takes place from the L1, which has a lower specific angular momentum. If the distance of L1 from the centre of mass of system is  $x_{\rm L1}a$, the specific angular momentum at L1 is 
$j_{\rm L1}=\Omega x^2_{\rm L1}a^2$. The factor  $f$ in equation (\ref{deff}) is then $x_{\rm L1}^2/x_{\rm p}^2$. Including a factor $1-\epsilon$ to parametrise anything that happens to the angular momentum after the stream leaves L1, equation (\ref{fdot}) becomes:
\beq  
{\dot a\over a}={2\over 1+q}{\dot M_{\rm p}\over M_{\rm p}}[{x_{\rm L1}^2\over x_{\rm p}^2}(1-\epsilon)+q^2-1].\label{fdot1}\qquad 
\eeq
Including the gravitational attraction of the planet on the stream leads to a small positive value of $\epsilon$ (next subsection). equation (\ref{fdot1}) with $\epsilon=0$ is our baseline for comparison with other parametrizations of the angular momentum exchange. We refer to it in the following as our `minimal assumption'.

The lower angular momentum of L1 compared with the planet's centre of mass has a significant effect, even for small mass ratios, see Fig.\ \ref{L1j}. This can be understood from the size of the planet's Roche lobe, which scales approximately as $q^{1/3}$. At a mass ratio $q=10^{-3}$, for example, the distance of the L1 (in uint of the orbital separation) from the planet's centre of mass is 100 times larger than the value of $q$ itself. As a result, the orbit tends to expand on mass loss, though not as much as under the conventional assumption of conservation of orbital angular momentum. This is illustrated further in section \ref{results} (see Fig.\ \ref{jp-m} and accompanying text).

\subsection{Stream-planet interaction}
\label{discplanet}

Metzger et al.\ (2012) include direct transfer of angular momentum from the orbit to the host, under the assumption that the accreted mass has the specific angular momentum of a Kepler orbit around the surface of the host, which is less than that of the orbit.  The assumption is  that the angular momentum still to be accounted for is transferred back to the orbit, through the formation of an accretion dics. Tidal interaction of the planet truncates the disc by removing angular momentum from its outer parts, returning it to the orbit. 
 
A comparison with CV dicss is informative. Paczy\'nski (1977) derived an upper limit to the truncation radius as the last non-intersection orbit, and finds that observed disc sizes in CVs are actually somewhat smaller. An extensively studied example is WZ Sge (mass ratio $q=M_2/M_1=0.08$). Spruit \& Rutten (1998) find a disc radius $r_{\rm d}/a\approx 0.37$, compared with a Paczy\'nski's maximum of about 0.57. For lower mass ratios a maximum to the disc radius is the 2:1 resonance radius (Lin \& Papaloizou 1979), which gives a limit of $r/a=2^{-2/3}=0.63$. In the case of a Roche lobe filling planet of $1\,M_{\rm J}$ with $1\,R_{\rm J}$ orbiting a host star of $1\,M_{\bigodot}$ with $1\,R_{\bigodot}$, the surface of the star is at $R_*= 0.45 a$ (cf.\ Fig. \ref {trajectory}).
 
Given that the observed truncation radii in low-mass ratio systems are rather smaller than the theoretical limit, there may not be much room  for a disc in the case of Roche lobe filling planets orbiting MS or larger host stars. In this case the accreting material should be expected to mix into the surface of the host star by shear instabilities, thereby transferring its angular momentum into the slowly rotating envelope instead of returning it to the orbit. In this case the only angular momentum returned to the orbit is related to the lower specific angular momentum of the stream leaving L1. 

Though most of the mass will be accreted on to the host star fairly soon, some of the stream may find itself kicked, or viscously spread, to a specific angular momentum outside that of the planet orbit. Much like in Type II migration of planets in discs, tidal interaction with the planet can keep such circumbinary material from accreting. In the course of the mass transfer process it will instead accumulate outside the planet orbit,  spreading outward and extracting orbital momentum from the planet. As in the case of circumbinary discs in CVs (Spruit \& Taam 2001), the fraction of mass that spills across the orbit instead of accreting does not have to be very large to cause substantial angular momentum loss in the long run [in the CV cases studied by Taam, Sandquist \& Dubus (2003): a fraction $\sim 10^{-4}$]. In this scenario mass loss from the planet will cause angular momentum to be {\it lost from} the orbit instead of being added to it. Though possibly important for secular evolution of the orbit, it is too slow to be relevant for the dynamical stability of mass transfer, however.

\subsection{Stream-magnetosphere interaction}

Much of the orbital evolution of close-in planets is likely to take place in late stages of the star's formation, when observed surface field strengths are a kilo-Gauss or more (e.g. Basri, Marcy \& Valenti 1992; Johns-Krull, Valenti \& Koresko 1999; Johns-Krull 2007; Yang \& Johns-Krull 2011). The radius of the magnetosphere: the boundary between the accretion disc and the stellar magnetic field, would then be a few stellar radii. The distance where Roche lobe overflow becomes relevant is well inside this distance. The mass-transferring stream is then strongly affected by the stellar magnetic field, certainly in the early phases of mass transfer, when the stream is still weak. In older systems the magnetospheric fields are much weaker, on the order of a few tens of G for ages $>$ 1 Gyr, the magnetic stresses factors $> 10^3$ smaller and probably negligible in this context.

As observed in the case of magnetic CVs (`polars', Krzeminski \& Serkowski 1977), the  interaction of the  magnetic field with the stream shreds it into an `accretion curtain'. The angular momentum of the stream is transferred to the star through magnetic stress in this curtain. The stream's angular momentum does not return to the planet: it is added to the spin angular momentum of the star. In terms of angular momentum transfer, the magnetosphere increases the effective radius of the star. The torque exerted in this way saturates when it reaches the maximum the magnetic field can sustain. This complication should be kept in mind, but is not included in the calculations below.

\subsection{Effect of the evolutionary state of the host star}

Much of the planet formation process, including its migration to an orbit close enough for mass transfer to become relevant, is believed to take place when the star itself is still in the process of contracting to the MS, and larger than its nominal MS radius. The larger radius increases the probability that a  stream transferring mass from the L1 is intercepted by the host star, in which case all the stream's angular momentum is lost to the star instead of returning to the planet orbit (see section \ref{streamimp}).

\subsection{Gravitational interaction}
\label{gravi}
After leaving L1, the stream moves ahead of the planet. The pull of gravity of the planet reduces the angular momentum of the stream, returning some of it to the planet. Fig.\ \ref {jorb} shows how this causes the stream to lose angular momentum to the planet as it starts moving ahead of the planet. The calculations were done by straightforward integration of the equation of motion of a test particle in a two-point-mass binary, at a relative numerical accuracy of $10^{\mathrm{-3}}$ or better (the code is described in Spruit 1998). For mass ratios of order 0.001, this effect is on the order of a few per cent. Continued interaction with planet and host star during its somewhat non-circular orbit causes additional smaller variations in its angular momentum with respect to the centre of mass of the system.

\section{Stability of mass transfer}
\label{eop}

The balance of the planet's angular momentum, whether positive or negative, depends on a number of factors. In the following we explore their effect on the fate of the planet: whether it survives intact before plunging into its host, or instead loses mass gradually by stable mass loss though L1 driven by tidal torques, or goes through unstable mass transfer, disrupting on a short, eventually dynamical time-scale. 

Some factors can be included with sufficient accuracy: the specific angular momentum of L1 (as opposed to the planet's centre of mass), the gravitational torque of the planet on the mass-transferring stream, the conditions for direct impact of the stream or that of the planet itself on the host star, and the role of the mass-radius relation of the planet. Disc spreading and its possible hydrodynamic interaction with the planet can only be included with additional assumptions. 

\begin{figure}
\center{\includegraphics [width= 0.47\textwidth]{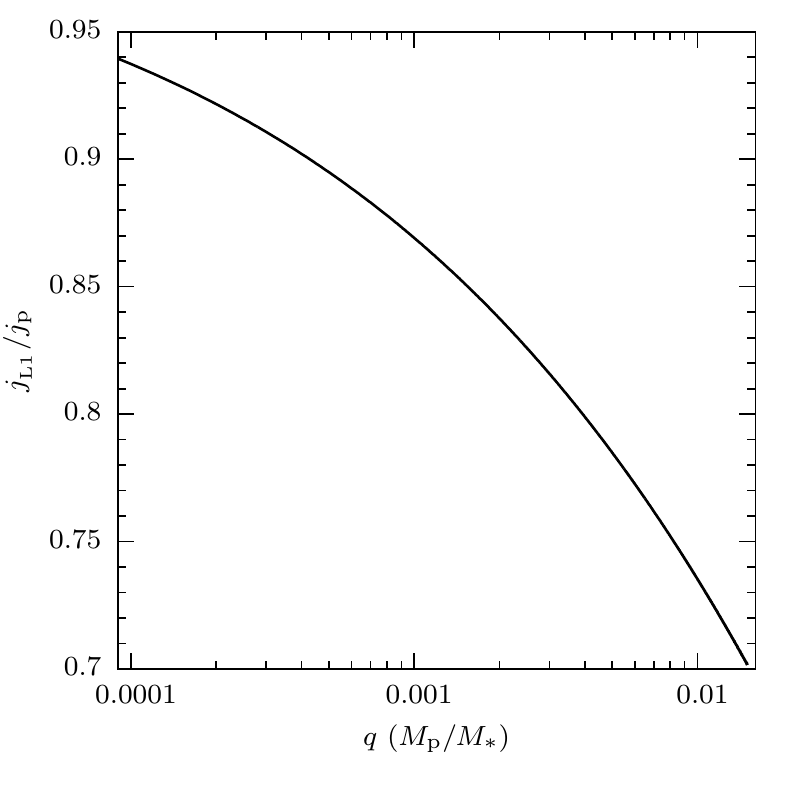}}
\caption{Specific angular momentum of the L1, measured with respect to the centre of mass of the system and normalized to the specific angular momentum of the planet's centre of mass, as a function of the mass ratio $q$.}
\label{L1j}
\end{figure}

\begin{figure}
\center{\includegraphics [width= 0.47\textwidth]{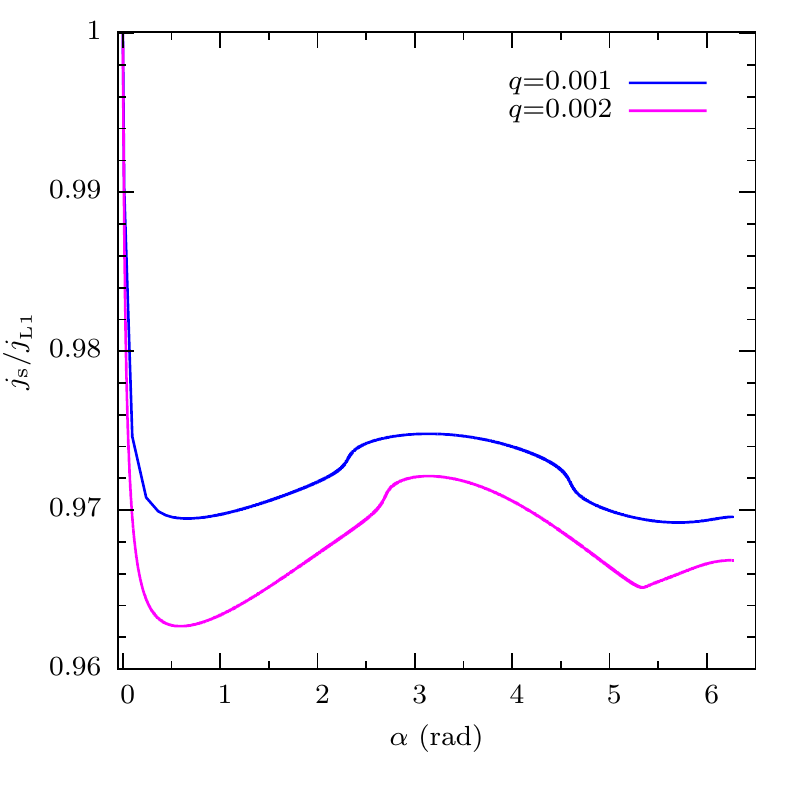}}
\caption{Specific angular momentum of the stream, measured with respect to the centre of mass of the system and  normalized to the specific angular momentum at the L1, as a function of orbital phase. Variation is due to the gravitational pull of the planet.}
\label{jorb}
\end{figure}

\subsection{Dynamical mass transfer}
\label{dyntrans}

\begin{figure}
\center{\includegraphics [width= 0.47\textwidth]{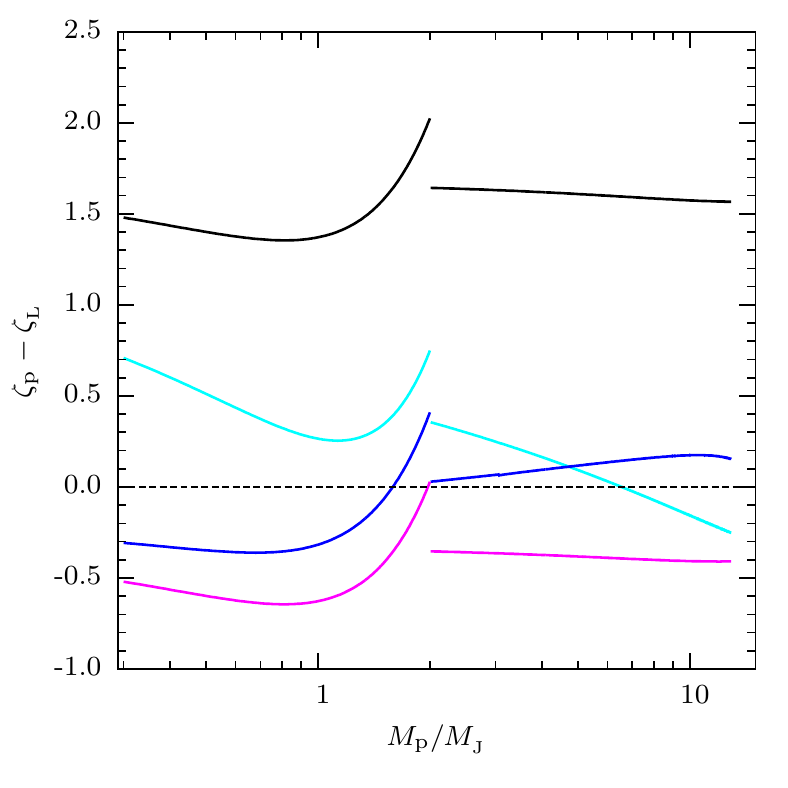}}
\caption{Dynamical instability of mass transfer under different assumptions for the return of angular momentum from the stream to the planet orbit. The Roche lobe filling gas giant planets orbit a Sun-like star (1.0 $M_{\mathrm{\bigodot}}$, 1.0 $R_{\bigodot}$). The planetary model (entropy ${\rm 9 \ k_{\rm B} \ {baryon}^{-1}}$, from Spiegel \& Burrows 2012) has different fits for the regions of $M_{\rm p}/M_{\rm J} \leq 2$ and $M_{\rm p}/M_{\rm J}>2$. The black dashed line shows $\zeta_{\rm p}=\zeta_{\rm L}$. Mass transfer is dynamically unstable below the dashed line. Black: all stream angular momentum assumed to be returned to the orbit. Cyan: same, except taking into account angular momentum lost by mass settling on the host star. Blue: stream carries the angular momentum of its own orbit rather than that of the centre of mass of the planet, but otherwise does not transfer angular momentum back to the planet orbit. Magenta: assuming mass transfer does not change specific angular momentum of the orbit.}
\label{zetas}
\end{figure}

\begin{figure}
\center{\includegraphics [width= 0.47\textwidth]{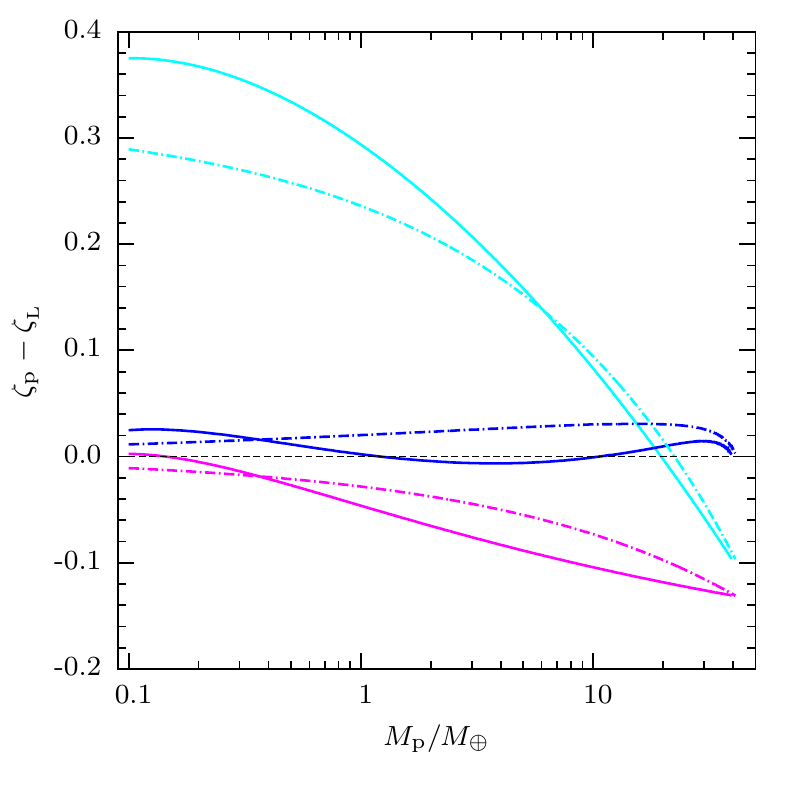}}
\caption{Same as Fig.\ \ref{zetas}, but for planetary models with pure silicate composition from Fortney et al. (2007) (solid curves), and Seager et al. (2007) (dot-dashed curves). The black curves are not shown here.}
\label{silicams}
\end{figure}

Secular angular momentum loss by tidal interaction (in the case of close binaries:  by magnetic braking) shrinks the Roche lobe of the planet until it touches the surface of the planet. The mass loss through L1 is then dynamically unstable if the radius of the planet $R_{\rm p}$ decreases less than the radius $R_{\rm L}$ of the Roche lobe. If $\zeta_{\rm p}={\rm d}\ln R_{\rm p}/{\rm d}\ln M_{\rm p}$ is the planet's adiabatic mass-radius exponent (cf. Ritter 1988), and  if $\zeta_{\rm L}={\rm d}\ln R_{\rm L}/{\rm d}\ln M_{\rm p}$ is the corresponding exponent of the Roche lobe radius $R_{\rm L}$ (Eggleton 1983), mass transfer is dynamically unstable when $\zeta_{\rm p}-\zeta_{\rm L}<0$. The planet's exponent depends on its internal structure. As in the case of low-mass MS stars, the radius of a gas giant planet increases when losing mass adiabatically:  $\zeta_{\rm p}<0$. The Roche exponent $\zeta_{\rm L}$, on the other hand, depends on what happens with the angular momentum carried by the mass-transferring stream, as discussed above. The consequences for different assumptions are shown in Fig.\ \ref{zetas}, for the example of hypothetical planets of Jupiter composition with entropy $ {\rm 9 \ k_{\rm B} \ baryon^{-1}}$ (Spiegel \& Burrows 2012) orbiting a $ 1 \ M_{\bigodot}$ star with $1 \ R_{\bigodot}$. [The prominent discontinuity at $2\, M_{\rm J}$ results from the use   of different fitting formulas on either side.]

If all angular momentum of the mass lost is returned to the orbit [black curve, $f=0$ in equation (\ref{fdot})] the mass transfer is dynamically stable, the orbit widens due to `angular momentum conservation'. 
 
As a second example, assuming that the mass transferred is accreted by the host star with the specific angular momentum of a Kepler orbit around its surface as is done in Metzger et al.\ (2012). This yields the cyan curve. An assumption made in this case is that the specific angular momentum $j_{\rm s}$ carried by the stream is that of (the centre of mass of) the planet, $j_{\rm p}$. Mass transfer is then dynamically unstable when the planet's mass is higher than 6.6. 

Mass loss is from L1, however, which has a lower specific angular momentum than the planet's centre of mass (Fig.\ \ref{L1j}). Mass loss from L1 therefore increases the mean specific angular momentum of the mass remaining in the planet. Taking this into account, but assuming that no further angular momentum feedback takes place yields our `minimal assumption' (the blue curve). This case applies if the stream hits the surface of the star directly instead of evolving into a disc (cf. section \ref{streamimp}), or if interaction between planet and disc is ignorable (see discussion in section \ref{discplanet}). Mass loss is now dynamically unstable for giant planetary masses up to about 1.6 $M_{\rm J}$ and stable above, up to the mass where direct merger occurs (section \ref {dmer}). The results are somewhat sensitive to the assumed internal structure of the planet, evident also from the break in the curves at $2 \ M_{\rm J}$ due to a discontinuity in the fits used in Spiegel \& Burrows (2012).

For comparison, the magenta curve shows the result of arbitrarily assuming that angular momentum exchange is such that the specific angular momentum of the planet does not change, i.e. ignoring that mass loss is from L1 instead of the planet's average ($f=1$ in equation (\ref{fdot})). Mass transfer is now unstable whenever the planet's adiabatic mass-radius exponent is negative, somewhat more unstable than the cyan and blue curves.

\begin{figure}
 \center{\includegraphics [width= 0.46\textwidth]{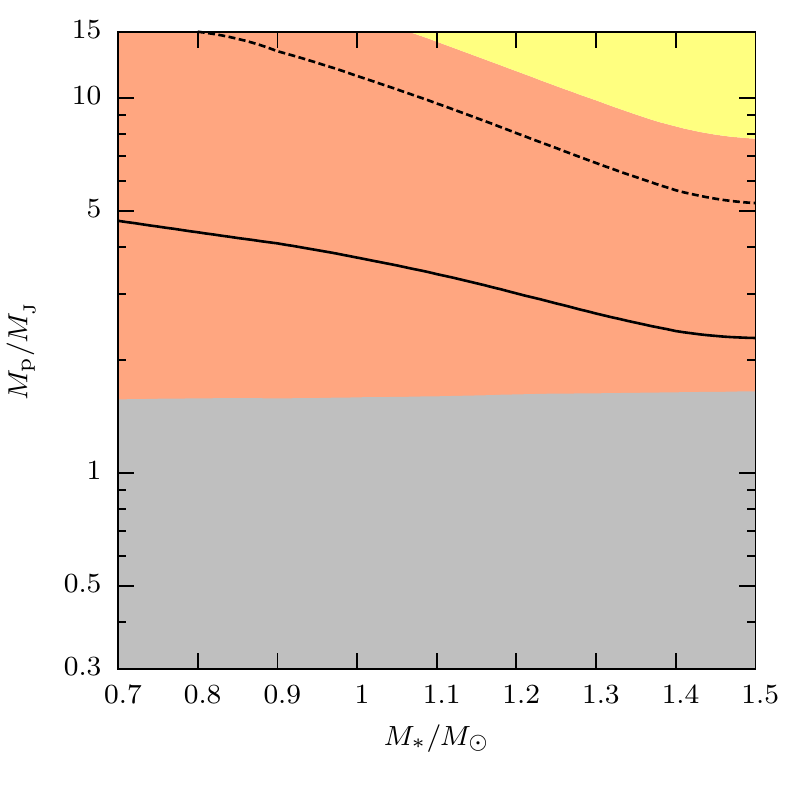}}
\caption{Regimes of mass transfer for gas giant planets orbiting zero-age main sequence stars. The `minimal assumption' is used as the angular momentum transfer model (see text). Yellow: the planet enters the star before overflowing its Roche lobe (direct merger). Orange (grey): dynamically stable (unstable) Roche lobe overflow. Above the black solid line the stream impacts the surface of the star. Above the black dashed line the star touches L1 before the planet does (see discussion in text). ZAMS mass-radius relation is from A.\ Weiss. Planet model is the same as in Fig. \ref{zetas}.
}
\label{zamsgiant}
\end{figure}

\begin{figure}
 \center{\includegraphics[width= 0.46\textwidth]{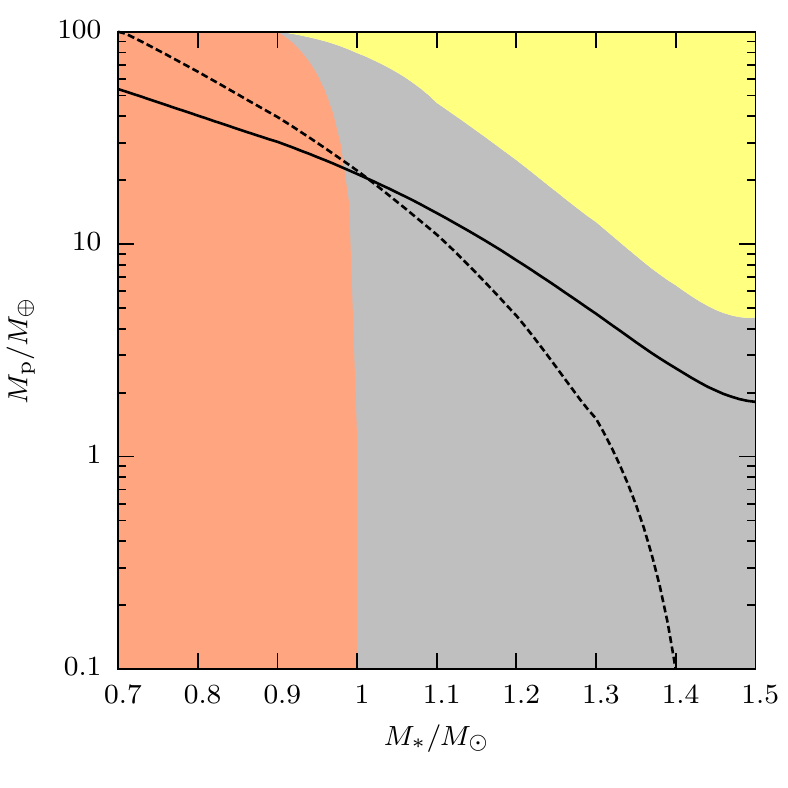}}
 \caption{Same as Fig. \ref{zamsgiant}, but for planetary model with pure silicate composition from Fortney et al. (2007).}
 \label{zamssilicate}
\end{figure}

Whether mass transfer is dynamically stable or not depends on the mass-radius relation of planet. Fig. \ref{silicams} shows the same as Fig.\ \ref{zetas} for planets of pure silicate composition, for two assumed equations of state (Fortney, Marley \& Barnes 2007; Seager et al. 2007). Except for the blue curve, the trends with planetary mass are the same as for gas giant planets. The blue curves (mass transferred has the angular momentum of the stream) are very close to the stability boundary ($\zeta_{\rm p}=\zeta_{\rm L}$). This reflects the fact that rocks are much less compressible than a gas, so their mass-radius exponent  ($R_{\rm p} \propto  {M_{\rm p}}^{\rm {1/3}}$) is similar to that  of the Roche lobe, especially in the lower mass range, in contrast with the negative exponent of isentropic gas spheres. Models with pure water or pure iron compositions (not shown) have mass-radius exponents qualitatively similar to those of silicate models. Fig. \ref{silicams}, shows that, for the `minimal assumption', the results are very sensitive to the planetary model. For the planetary model of Fortney et al.(2007)  mass transfer is unstable for initial masses in the range $1 \ M_{\bigoplus} < M_{\mathrm{p}} < 10 \ M_{\bigoplus}$; for the planetary model of Seager et al.(2007) mass transfer is marginally stable in this mass range. In the following, the planetary models of Fortney et al.(2007) are used.

In addition to the planet's mass and its adiabatic mass-radius relation, the boundary between dynamically stable and unstable mass transfer depends on the stellar mass-radius relation. Fig. \ref{zamsgiant} shows this dependence for gas giant planet as a function of mass of zero-age main sequence (ZAMS) star. The angular momentum feedback model is the `minimal assumption'. The orange region and grey region of Fig. \ref{zamsgiant} denote the dynamically stable and unstable mass transfer, respectively.

\begin{figure}
\center{\includegraphics[width= 0.52\textwidth]{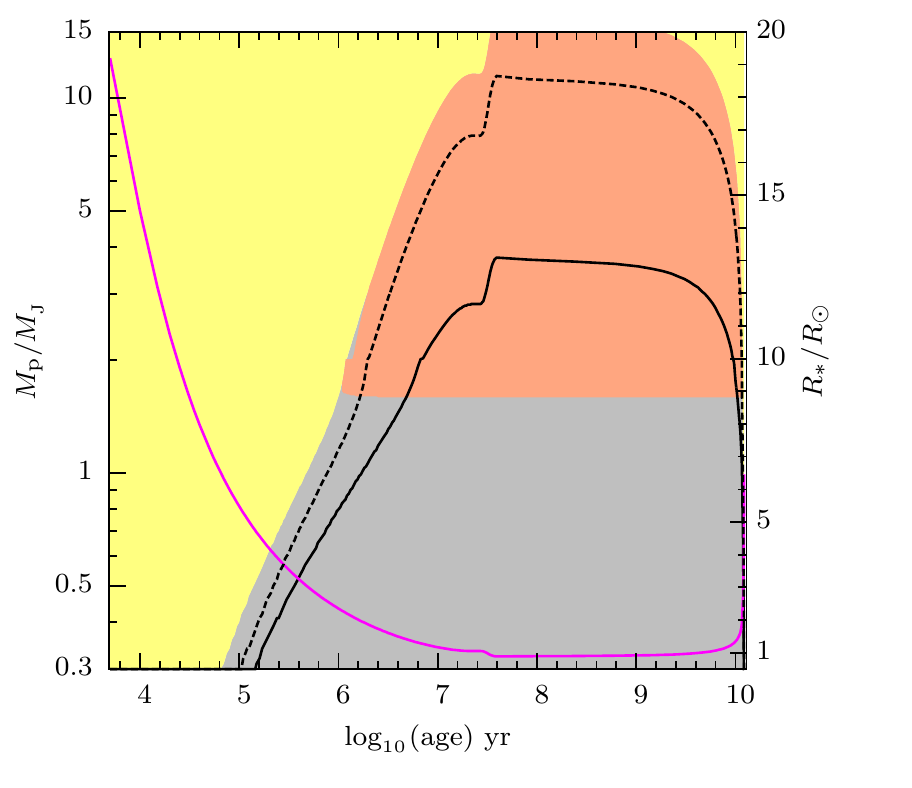}}
 \caption{Regimes shown in Fig.\ \ref{zamsgiant} as a function of age, for a host star of $1\ M_{\bigodot}$. Violet line: radius of the star (scale on the right).}
 \label{solargiant}
\end{figure}

\begin{figure}
 \center{\includegraphics[width= 0.52\textwidth]{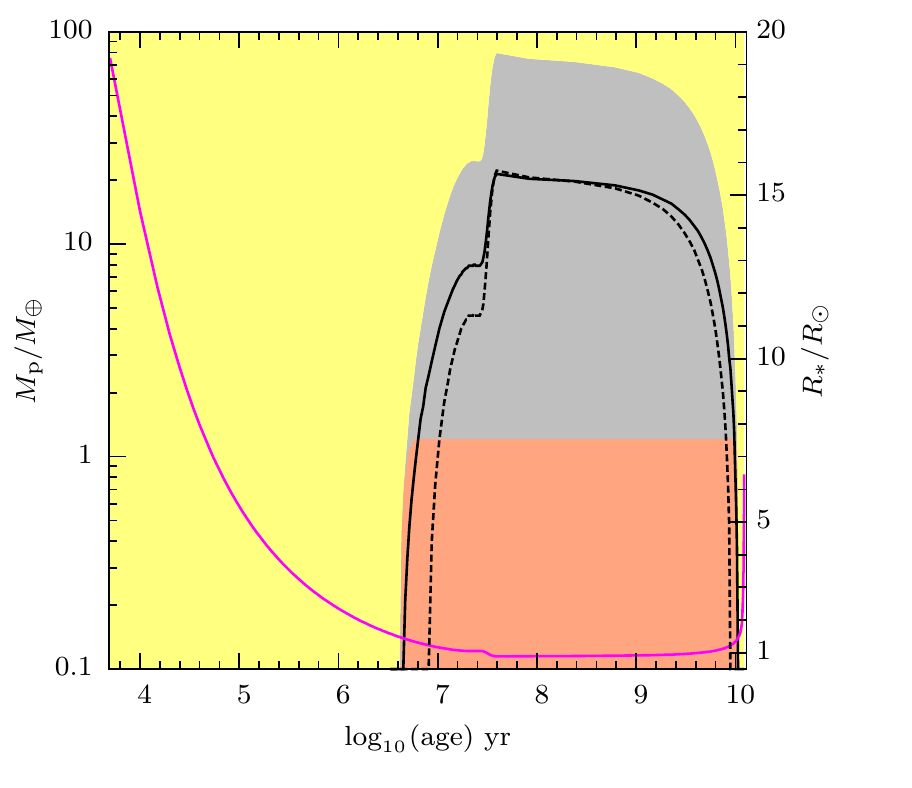}}
 \caption{Same as Fig.\ \ref{solargiant}, but for planetary model with pure silicate composition from Fortney et al. (2007). Violet line: radius of the star (scale on the right).}
 \label{solarsilicate}
\end{figure}

\begin{figure}
\center{\includegraphics[width= 0.52\textwidth]{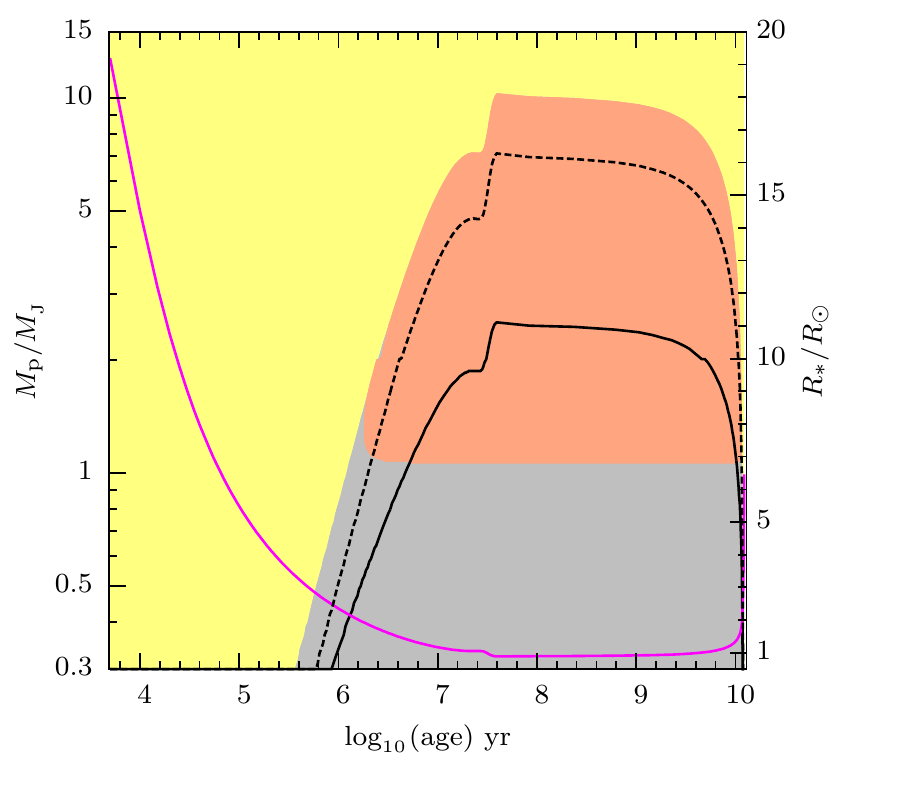}}
\center{\includegraphics[width= 0.52\textwidth]{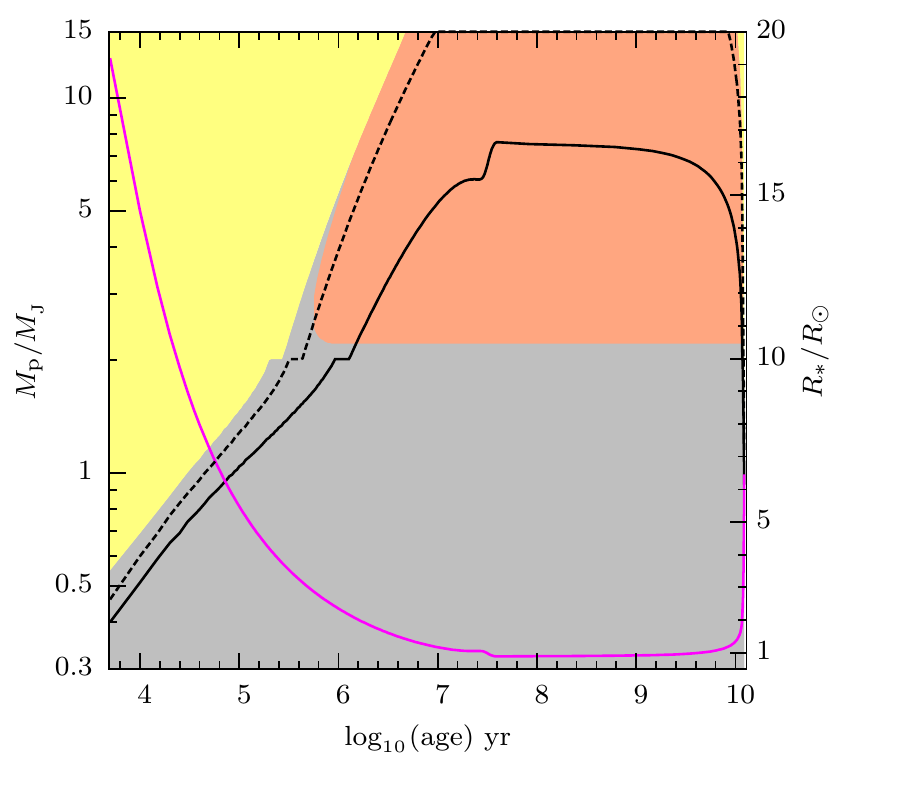}}
\caption{Same as Fig. \ref{solargiant}, but for gas giant planetary models with entropy $7 \ k_{\mathrm{B}} \ \mathrm{baryon}^{-1}$ (top panel), and $11 \ k_{\mathrm{B}} \ \mathrm{baryon}^{-1}$ (bottom panel). Violet line: radius of the star (scale on the right). The kinks at 2 $M_{\mathrm{J}}$ reflect the discontinuity in the fitting formulas of Spiegel \& Burrows (2012).}
 \label{solargiants7}
\end{figure}

Fig. \ref{zamssilicate} is the same as Fig. \ref{zamsgiant}, but for planets of pure silicate composition. The change from a horizontal to vertical boundary between Fig. \ref{zamsgiant} and Fig. \ref{zamssilicate} reflects the different signs of the mass-radius exponent between giant planet and rocky planet. Fig. \ref{solargiant} shows the dependence for gas giant planet as a function of stellar age (1 $M_{\bigodot}$) during the pre-main sequence (PMS) stage and the MS stage. Note that the entropy of the gas giant planets are kept constant in our calculations, on the argument that the cooling time is much longer than the mass transfer instability time-scale. There is a tiny grey area (around $\mathrm{{10}^{6.2}}$ yr and 2 $M_{\mathrm{J}}$) nearly invisible in the Fig. \ref{solargiant} resulting from the discontinuity in the fits used in Spiegel \& Burrows (2012). The parts of the dashed and the solid lines in Figs.\ \ref{zamsgiant}-\ref{solargiants7} at around $\mathrm{10^{7.5}}$ yr are not smooth due to the variation of star's radius during the deuterium burning. Fig. \ref{solarsilicate} shows the same as Fig. \ref{solargiant} for planets of pure silicate composition. The stellar models used here were provided by A. Weiss (cf. Weiss \& Schlattl 2008).

Below the yellow direct merger region is a narrow zone where the host star fills its Roche lobe {\em before} the planet does (above the dotted line in Figs \ref{zamsgiant}--\ref{solargiants7}). This does not mean that the star actually transfers mass to the planet, however, since the star does not corotate with the orbit. The effective potential of which L1 is a saddle point applies only for uniform rotation of the whole system. If the star is not corotating,  its surface does not sense the centrifugal force that is assumed in the calculation of its Roche lobe. The dotted line in the figures is therefore shown mostly for curiosity.

The dependence of the results on the thermal initial condition of the planet is illustrated in Figs. \ref{solargiant} and \ref{solargiants7}. It shows how the region of unstable mass transfer increases in size between entropy values of ($7 \ k_{\mathrm{B}} \ \mathrm{baryon}^{-1}$) and ($11 \ k_{\mathrm{B}} \ \mathrm{baryon}^{-1}$). Direct merger of the planet with the star (see section \ref{dmer}) occurs more easily for planet with low entropy than that with high entropy (larger yellow region), because a planet with low entropy is denser than with high entropy. The unstable grey area has a narrow upward extension around $10^6$ yr. This is related to the inclusion of the gravitational torque of the planet on the stream path (cf. Fig. \ref{jorb}).

\subsection{Stream impact}
\label{streamimp}

\begin{figure}
\includegraphics [width= 0.47\textwidth]{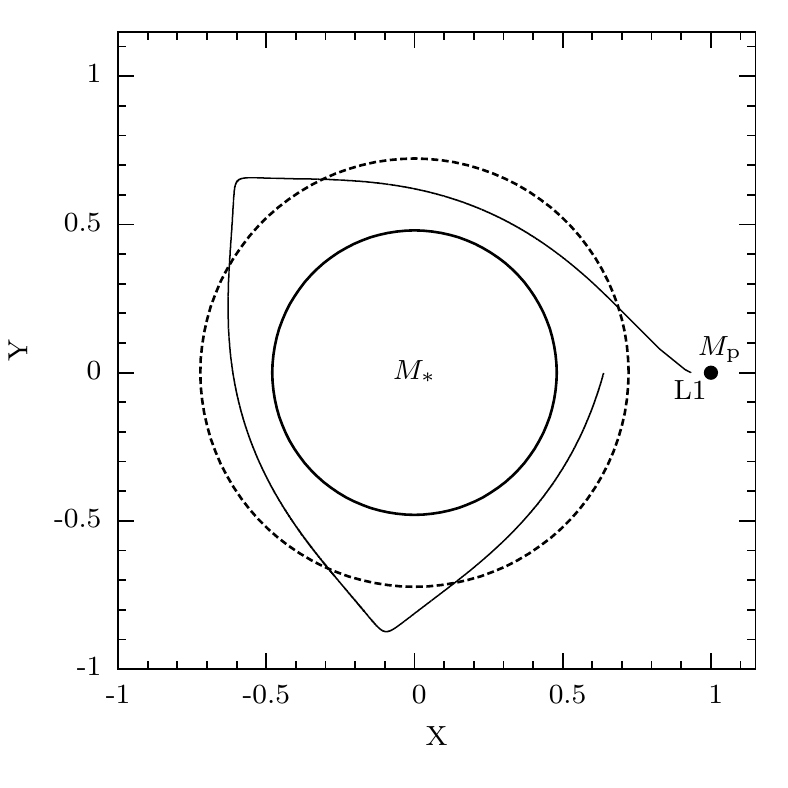}
\caption{Trajectory of the stream (thin black line) from a Roche lobe filling $1 \ M_{\rm J}$ planet with $1 \ R_{\rm J}$ around a $1 \ M_{\bigodot}$ star. Inner solid circle: radius of the present Sun. In this case, the stream travels around the host star in one orbital period without touching the surface of host star. Outer dashed circle: the radius of the host star at 3.94 Myr. With the larger stellar radius, 1.5 $R_{\bigodot}$, the stream impacts the stellar surface soon after leaving L1.}
\label{trajectory}
\end{figure}

\begin{figure}
\includegraphics [width= 0.47\textwidth]{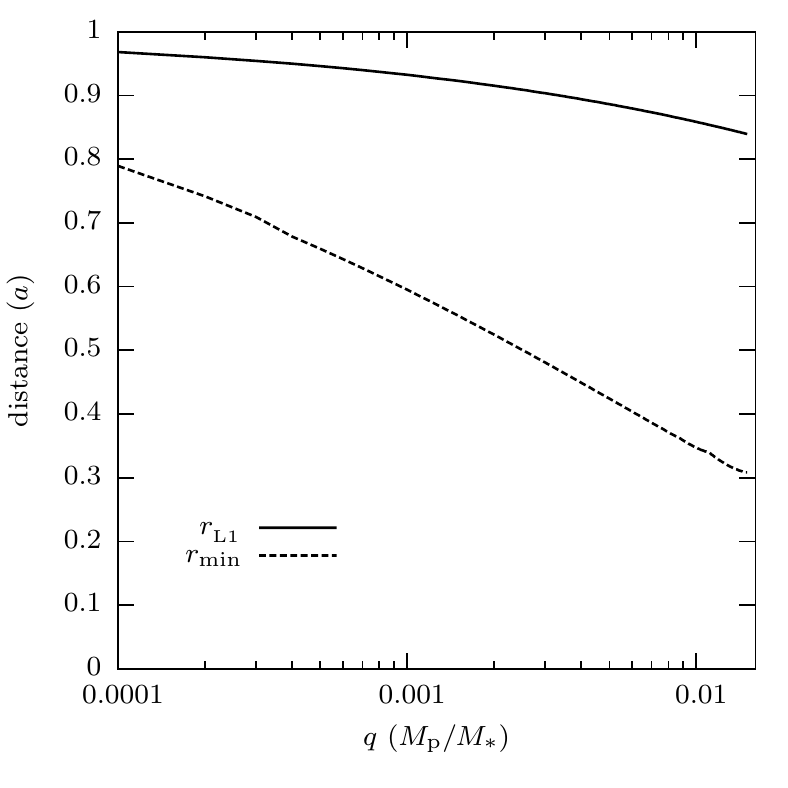}
\caption{Distances from the centre of the host star, as a function of mass ratio. solid line: location of L1, dashed line: minimum distance of a stream leaving L1.}
\label{rmin}
\end{figure}

Depending on mass ratio and size of the host star, the stream leaving L1 can hit the surface of the star instead of forming a disc. When the accretor is a MS star, as in the case of Algol type binaries, this happens in a wide range of mass ratios. It can happen also at the low mass ratios of mass-transferring planets. This is illustrated in Figs \ref{zamsgiant}--\ref{rmin}. The region between the yellow area and the black solid lines of Figs \ref{zamsgiant}--\ref{solargiants7} equivalently shows that stream impact will occur in terms of the planet's mass needed for its stream to intersect the stellar surface. Fig.\ \ref{trajectory} shows the orbit of the stream leaving the L1 of a $1 \ M_{\rm J}$ gas giant with $1 \ R_{\rm J}$ around a $1 \ M_{\bigodot}$ star of two different ages. Fig.\ \ref{rmin} shows the minimum distance of the stream orbits to the centre of the star, as a function of the mass ratio $q$. The calculations of Figs \ref{trajectory} and \ref{rmin} were done with the code in Spruit (1998) as in Fig. \ref{jorb}.

At the size of the current Sun the stream does not intersect the stellar surface (Fig.\ \ref{trajectory}). At an age of 3.94 Myr, when the host star was 50 per cent larger, the stream from a mass-transferring planet (1 $M_{\mathrm{J}}$) would have intersected the stellar surface, and the stream's angular momentum would have been transferred to the star. The `minimal assumption' for angular momentum return would then apply (cf.\ Fig.\ \ref{zetas} and associated text in section \ref{dyntrans}).

The dashed lines of Fig.\ \ref{tin1} (gas giant planet) and Fig.\ \ref{tin2} (planet with pure silicate composition) show what this means for a PMS star, as a function of planetary mass. The time $t_{\rm in}$ since the star's birth during which its radius is large enough to intercept the stream increases with increasing mass ratio. The stream from a hypothetical planet of 1 $M_{\rm J}$ filling its Roche lobe, for example, would be intercepted for the first 6.2 Myr of the PMS life of a star with $1.3 \ M_{\bigodot}$, or about 85 per cent of its contraction time to the MS. 

\subsection{Direct merger}
\label{dmer}
\begin{figure}
\includegraphics [width= 0.47\textwidth]{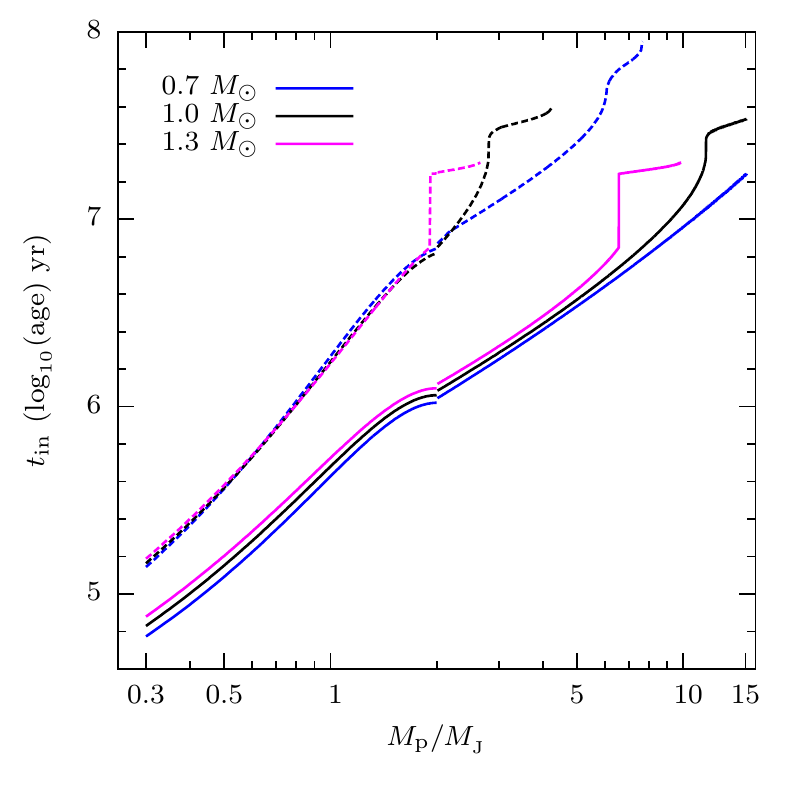}
\caption{Minimum planetary mass $M_{\rm p,min}$ needed for stream impact (dashed lines) and direct merger of the planet (solid) as a functions of planetary mass and age of host stars of 0.7 $M_{\bigodot}$ (magenta); 1.0 $M_{\bigodot}$ (black); 1.3 $M_{\bigodot}$ (blue). The ends of the lines are not smooth because of the variation of the star's radius during deuterium burning. Planet model is same as in Fig. \ref{zetas}.}
\label{tin1}
\end{figure}

\begin{figure}
\includegraphics [width= 0.47\textwidth]{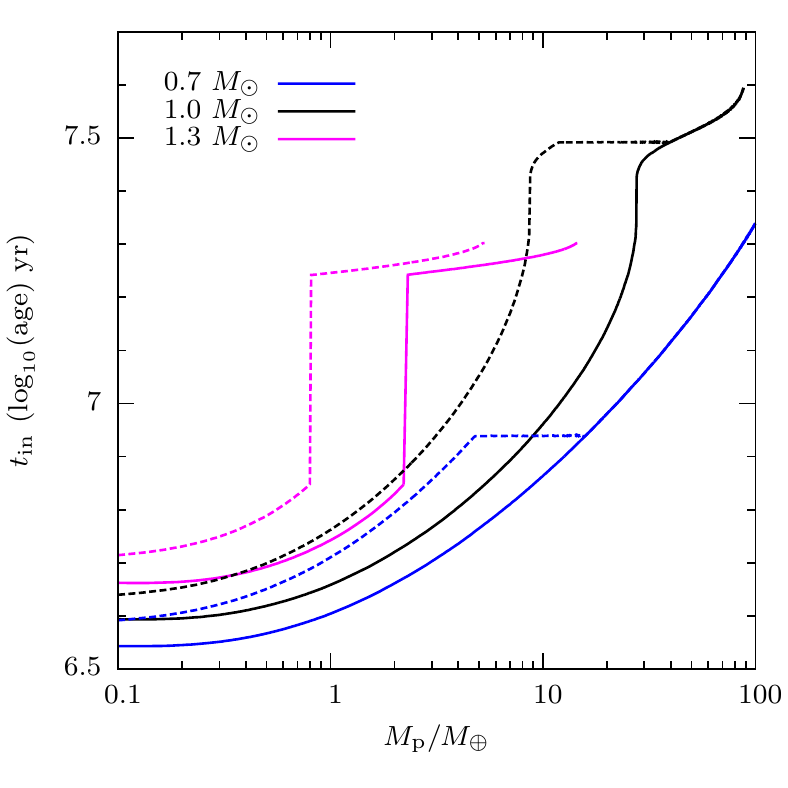}
\caption{Same as Fig.\ \ref{tin1}, but for planetary model with pure silicate composition from Fortney et al. (2007).}
\label{tin2}
\end{figure}

\ehs
Under more extreme conditions the planet itself (or rather: its L1) can reach the surface of its host before it fills its Roche lobe. In MS stars the density increases so quickly with depth below the photosphere that aerodynamic drag will cause spiral-in into the star almost immediately when this point in its orbital evolution is reached. Since the mean density of the star increases during its PMS evolution, it is more likely to happen to planets that already migrated close enough to their hosts during PMS contraction. The region in the $M_{\rm p}-M_*$ space where direct merger of the planet with the star occurs is shown by the yellow regions in Figs.\ \ref{zamsgiant}-\ref{solargiants7}. Figs. \ref{solargiant}, \ref{solarsilicate}, and \ref{solargiants7} show how direct merger of the system is more likely to happen during PMS contraction than during of MS. The solid lines of Fig.\ \ref{tin1} and Fig.\ \ref{tin2} show the minimum planetary mass for direct merger to happen, as a function of planetary mass and age of the host star. On the whole, direct merger is more likely to occur to planet-star systems containing a compact planet, a younger or a more massive host star.

\section{Time evolution of unstable mass transfer}
\label{dymt}

In the following we investigate the time evolution of the planet's orbit and mass in the case of dynamically unstable transfer (cf.\ Fig.\ \ref{zetas}). As discussed in the previous section, whether mass transfer is dynamically stable or not depends strongly about the assumptions made about the fate of the angular momentum carried by the stream. This has led to contrasting estimates in the literature for the occurrence of dynamical mass transfer. 

The presence of a surface (the host star) close to the stream  increases the likelihood of angular momentum being lost from the orbit directly to the star (section \ref{dyntrans}). Instead of angular momentum returning to the orbit through tidal interaction with a disc, as in the process governing the orbital evolution in compact-object binaries, mass transfer from the companion need not lead to widening of the orbit. The large moment of inertia and low rotation speed of the star make it a very effective sink of angular momentum. The small mass ratio of the planet-star system also causes the stream to stay much closer to the planet orbit  than is the case in CVs and X-ray binaries, increasing the likelihood that some of the stream spreads beyond the planet orbit into a circumbinary disc (see Spruit \& Taam 2001 for cases where this may also play a role in CVs). Like in the case of type-II migration of planets, the resulting torque on the orbit would cause it to shrink instead of expanding.

In the calculations following below the smaller specific angular momentum of the L1 and the gravitational pull of the planet  on the stream are taken into account, but no further angular momentum exchange with the orbit is included [our `minimal assumption', the blue curve in Fig.\ \ref{zetas}, equation (\ref{fdot1}) with $\epsilon$ as explained in section \ref{gravi}]. We consider this a representative case, possibly also the most likely.

\subsection{Calculations}
\label{calcs}

The locations of L1 and outer Lagrangian point (L2) follow from the Roche geometry, for which the expansions in Kopal (1959) were found accurate enough. These formulae and those by Eggleton (1983) are not quite correct especially for rocky or incompressible fluid planets, since these have their mass concentrated more towards the surface than gaseous planets, for which a point mass approximation is adequate for the Roche potential. We have not attempted to correct for this, since it will just shift the orbital separation where the planet fills its Roche lobe less than 20 per cent in orbital separation compared with the approximation of Roche radius for an incompressible fluid from Rappaport et al. (2013). We expect that this does not have a qualitative effect on the instability boundary in the $M_*-M_{\mathrm{p}}$ plane.

\begin{figure*}
\begin{center}
\includegraphics [width= 0.95\textwidth]{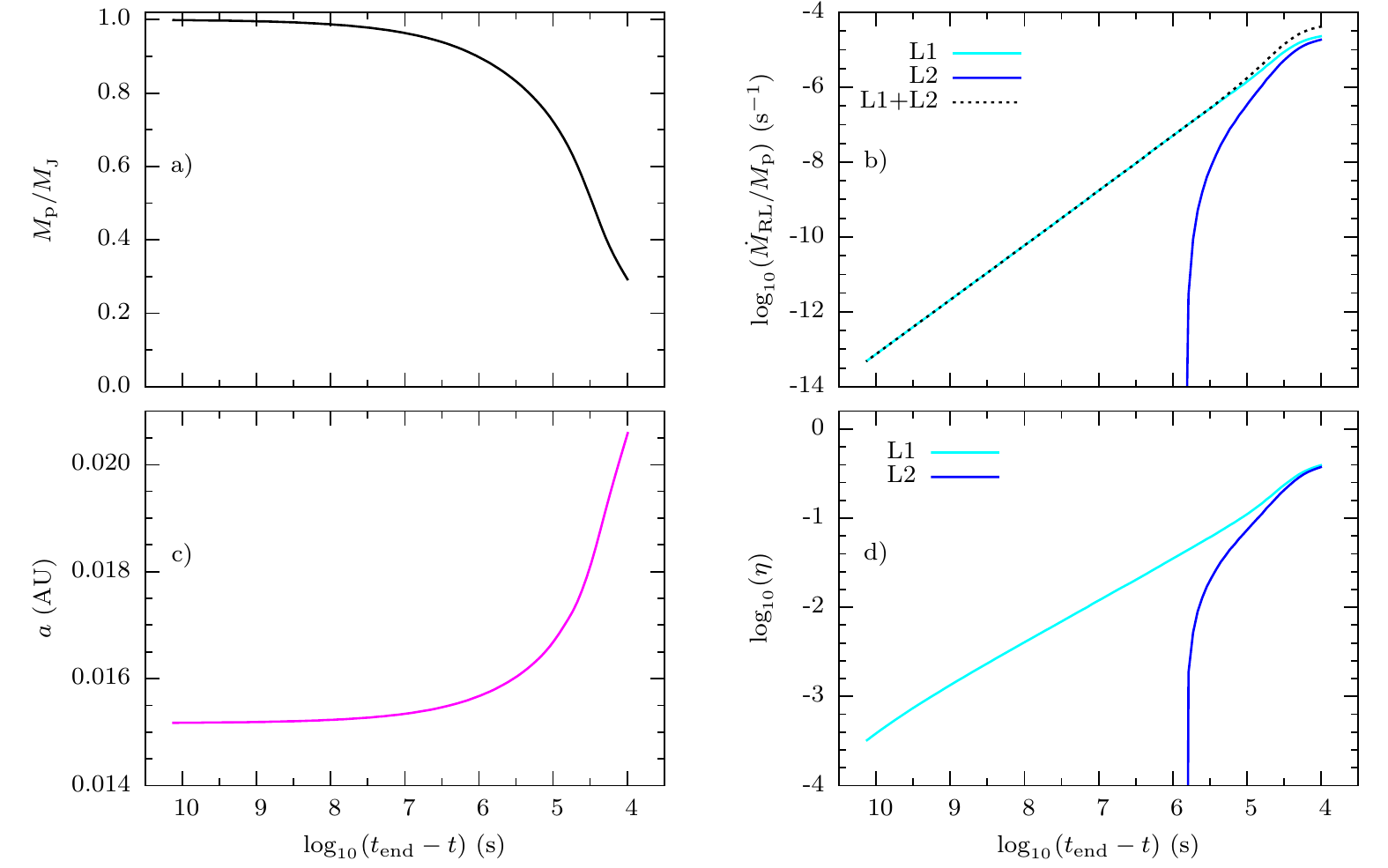}
\caption{Example of the evolution under dynamically unstable mass transfer of a planet of initial mass 1.0 $M_{\mathrm{J}}$ and radius 1.5 $R_{\mathrm{J}}$, orbiting a sun-like star (1.0 $M_{\bigodot}$, 1.0 $R_{\bigodot}$). Horizontal axis is the time left before final dissolution of the planet. Panel a: the planetary mass. Panel b: mass transfer rate from L1 (solid cyan line), L2 (solid blue line) and the total (L1$+$L2, black dotted line). Panel c: the orbital separation $a$. Panel d: overfill fractions $\eta$ at L1 (solid cyan line) and L2 (solid blue line).}
\label{mtgiant} 
\end{center}
\end{figure*}

Figure \ref{jorb} shows the trajectory of the stream in a corotating frame and the variation of $j_{\mathrm{s}}$ over one orbital period for mass ratios $q$ of 0.001 and 0.002. The wiggles in the curves show that at these low mass ratios, the gravity of planet and the displacement of the host star relative to the centre of mass of the system still affect the angular momentum of the stream to some degree. The maximum difference between $j_{\mathrm{s}}$ and $j_{\mathrm{p}}$ is about 16 per cent for $q=0.001$. This difference between specific angular momentum of the stream and that of the planet's centre of mass is taken into account in the calculations that follow.

\begin{figure}
\begin{center}
\includegraphics [width=0.47\textwidth]{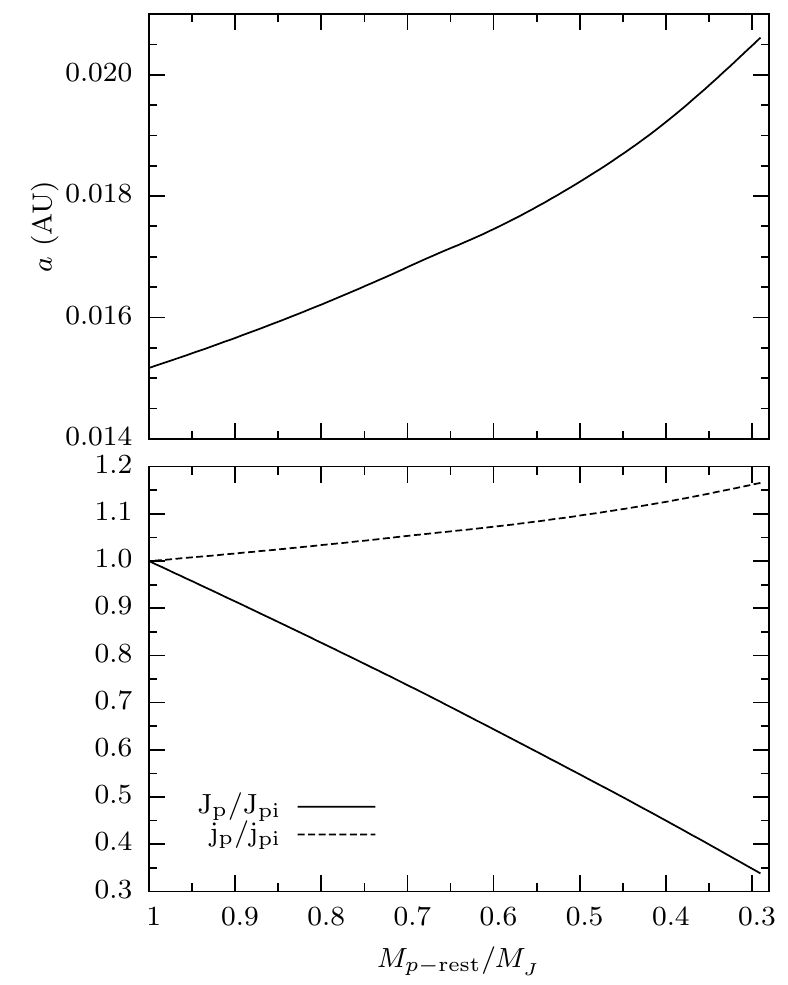}
\caption{The orbital evolution and angular momentum evolution of the planet shown in Fig. \ref{mtgiant}. Horizontal axis is the remaining mass of the planet ($M_{\mathrm{p-rest}}$). Top panel: the orbital separation $a$ as a function of $M_{\mathrm{p-rest}}$. Bottom panel, solid: orbital angular momentum of the planet ($J_p$) as a function of $M_{\mathrm{p-rest}}$, dashed: specific orbital angular momentum of the planet ($j_p$) as a function of $M_{\mathrm{p-rest}}$.}
\label{jp-m}
\end{center}
\end{figure}

\subsection{Mass loss model}
\label{lossmodel}
During mass transfer, the planet's radius exceeds the Roche lobe by an (initially small) amount. Calculation of the evolution of the planet's mass needs two elements: a model for the mass transfer rate as a function of this radius excess, and a model for the reaction of the planet's radius upon mass loss. The mass transfer rate is very sensitive to the difference between the radius of planet and the Roche Lobe radius, $(R_{\mathrm{p}}-R_{\mathrm{Roche}})/R_{\mathrm{p}} \equiv \eta$. The planet does not expand instantaneously, it takes the planet's dynamical time-scale, of order of the sound crossing time through its interior (Paczy{\'n}ski 1965; Paczy{\'n}ski, Zi{\'o}lkowski \& Zytkow 1969). For most of the time, this time-scale is short compared with the time-scale on which mass transfer evolves, so the resulting delay in the planet's response can be ignored. The calculation is stopped when the evolution time-scale has become comparable to the planet's dynamical time-scale.

\begin{figure}
\begin{center}
\includegraphics [width=0.47\textwidth]{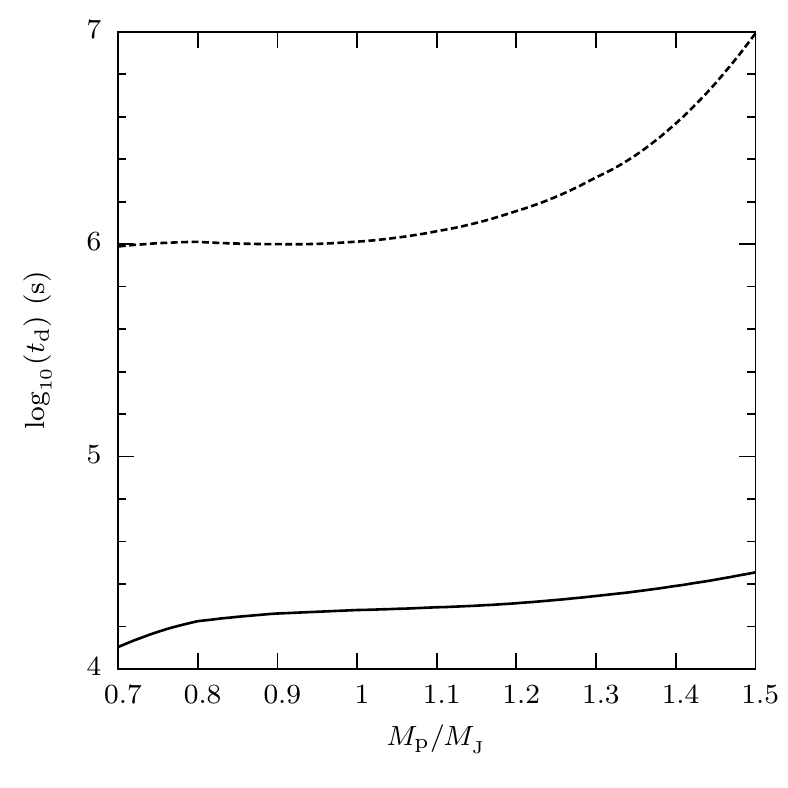}
\caption{Duration $t_{\mathrm{d}}$ of the transfer of the last 50 per cent (solid line) and 90 per cent (dashed line) of the mass of a gas giant planet, as a function of initial planetary mass. Host star is a Sun-like star (1.0 $M_{\mathrm{\bigodot}}$, 1.0 $R_{\bigodot}$).}
\label{fmass}
\end{center}
\end{figure}

\begin{figure*}
 \begin{center}
  \includegraphics[width= 0.95\textwidth]{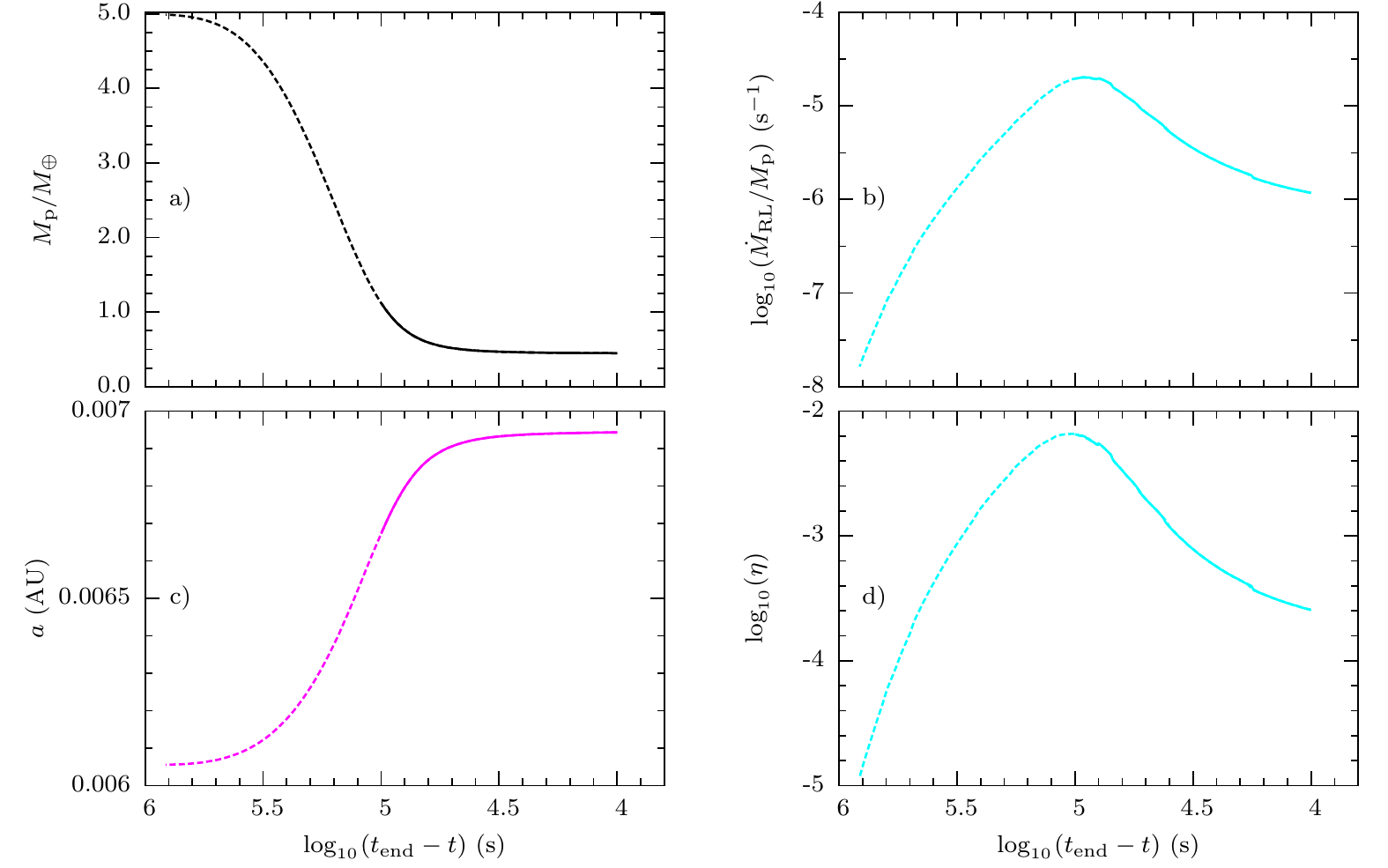}
\caption{Same as Fig. \ref{mtgiant}, but for a pure silicate planet. The initial mass and the radius of the planet are 5 $M_{\bigoplus}$ and 1.7 $R_{\bigoplus}$.}
   \label{mtsilicate}
 \end{center}
\end{figure*}

The mass flux depends on the temperature and density at L1: a model for the stratification of the planet's outer layers is needed for this. Since the duration of the dynamical mass transfer process is still short compared to the thermal evolution time of the planet interior, the expansion of the planet can be assumed to take place adiabatically. The mass evolution of the planet during mass transfer is obtained as the model of Savonije (1978, Appendix I), with the mass transfer rate computed as a function of the radius excess as in his equation (A10). In this model the stratification of the outer layers of planet is parameterized approximately with a polytropic relation $P = K \rho^{\mathrm{(1+1/n)}} $, with index $n$ and constant $K$, where $n = 1.73$, $K=1.34 \times 10^{12}$, determined by fitting to the corresponding equation of state  (with entropy $9 \ k_{\mathrm{B}} / \mathrm{baryon}$) from Saumon, Chabrier \& van Horn (1995). We assume that the parameters ($n$ and $K$) of the polytropic relation are constant during the planet's adiabatic expansion. The locations of L1 and L2 are taken from Kopal's (1959) approximation to the Roche geometry. For the density and temperature at these points the planet and Roche lobe are approximated as spheres centreed on the centre of mass of the planet. 

Mass transfer starts at L1. Mass loss through L2 is also included. At very low mass ratio $q$, mass loss starts simultaneously through both (e.g. simulation in Leinhardt et al. 2012). For the mass ratios considered here, $q\sim 10^{-3}$, mass loss through L2 is small at first but contributes significantly towards the end. The volume radius of Roche lobe of L1 ($R_{\mathrm{L1}}$) is taken from the approximate expression of Eggleton (1983). For mass loss through L2, an expression is needed for the planet's radius relative to the location of L2. In the spirit of the approximations used, we assume that the planet's radius $R_{\mathrm L2}$ at this point is given by
\begin{equation}
 R_{\mathrm{L2}} = R_{\mathrm{L1}} \frac{r_{\mathrm{L2}}}{r_{\mathrm{L1}}},
\end{equation}
where $r_{\mathrm{L1}}$ is the distance of L1 from the planet's centre of mass, $r_{\mathrm{L2}}$ is the distance of L2 from the planet's centre of mass.

\subsection{Results}
\label{results}

The orbital evolution model is as discussed in section\ \ref{dymt} and the equations are solved by a straightforward fourth-order Runge-Kutta scheme. As an example, we consider a gas giant planet and a Sun-like star (1 $M_{\bigodot}$, 1 $R_{\bigodot}$), and the `minimal assumption' case for the angular momentum exchange. The mass-radius relation of the planet is as shown in Fig.\ \ref{zetas}. The initial mass and radius of the planet are 1.0 $M_{\mathrm{J}}$ and 1.5 $R_{\mathrm{J}}$, respectively. The calculation starts when the planet just fills Roche lobe with a initial $\eta$ $\simeq$ 0.0003. The evolution of this specific planet-star system is shown in Fig.\ \ref{mtgiant}. As expected, the mass transfer initially develops quite slowly, while the mass of planet $M_{\mathrm{p}}$ and the orbital separation $a$ stay almost constant (panels a and c of Fig.\ \ref{mtgiant}). The mass transfer rate increases roughly a $(t_{\rm end}-t)^{-1.5}$ (panel b of Fig. \ref{mtgiant}), where $t_{\mathrm{end}}$ is the duration of mass transfer process for this specific planet-star system. At $(t_{\rm end}-t) \approx 10^6$ s the planet has lost 10 per cent of its mass, all of it through L1, while at $(t_{\rm end}-t) \approx 10^5$ s mass loss through L2 also becomes significant. The last 50 per cent of the planet's mass is lost on a time-scale of the order of several orbital periods ($\sim 10^{\mathrm{4}}$ s). Although the mass transfer rates through L2 is significant in the final stage, mass loss from L1  still dominates. Mass loss through L2 is more likely to feed into a circumbinary disc than loss through L1, and tidal interaction with this disc will be a source of angular momentum loss. This interaction would cause mass transfer to increase more dramatically when transfer through L2 sets in, but by the assumptions made this is not included in the model.

The lower left panel in Fig.\ \ref{mtgiant} shows how the orbit of (the remnant of) the planet expands during mass transfer. This happens even though in our `minimal assumption' there are no torques acting on the planet. It results from the fact that the specific angular momentum of the mass lost through L1 is less than average. As a result, the specific angular momentum of the remaining mass of the planet increases, moving the centre of mass outward. To illustrate this, Fig.\ \ref{jp-m} shows the evolution of the orbit, the planet's angular momentum (lower panel, solid) and its specific angular momentum (dotted). Angular momentum decreases as the orbit expands while specific angular momentum increases (somewhat). One can visualize this conceptually by imagining the mass transfer as a 2 step process: in the first, mass residing near L1 is lost. This moves the centre of mass of the remainder out a bit, equivalent to the increase of the mean specific angular momentum. In the second step the mass redistributes itself in the Roche lobe under conservation of mass and angular momentum. This process take place continuously. The fact that it works even for low mass ratios is a consequence of the weak dependence of Roche lobe size on mass ratio ($\sim q^{1/3}$, cf.\ section \ref{orbtohost}). With this dependence on $q$, the expansion factor converges to a finite value for $M_{\rm p}\rightarrow 0$, in contrast with the assumption of conservation of orbital angular momentum, by which the orbit would expand as $M_{\rm p}^{-2}$. 

Fig.\ \ref{fmass} shows the time spent for the last 90 per cent and the last 50 per cent of a gas giant planet's mass to be lost as a function of the initial mass. Somewhat independent of initial mass, the last 90 per cent of planetary mass is lost on a time-scale of the order of $10^{6}$ s ($\sim$ month), while the last 50 per cent is lost on a time-scale of the order of orbital period ($\sim 10^{4}$ s). 

Because of their different mass-radius relation, the evolution of silicate planets differs from the gas giant planets, especially for the `minimal assumption' case for angular momentum exchange. Fig. \ref{mtsilicate} shows results for the example of a pure silicate planet ($5 \ M_{\bigoplus}, 1.7 \ R_{\bigoplus}$, mass-radius relation from Fortney et al. 2007) orbiting a Sun-like star ($1 \ M_{\bigodot}, 1 \ R_{\bigodot}$). This mass-radius relation is based on the \textsc{aneos} equation of state (Thompson 1990). For the mass loss prescription needed (section \ref{lossmodel}) a polytropic fit  $P=K (\rho-\rho_{\mathrm{0}})^{\mathrm{(1+1/n)}}$, is made to this equation of state, yielding $n=1.19$, $K=2.98 \times 10^{11}$, and $\rho_{0} = 3.2 \ g/\mathrm{cm^{3}}$ is the density of rock (olivine) at low pressure. The mass transfer rate is high because of the high surface density of a rocky planet. The transfer process starts unstable, but the transfer rate starts to decrease when the planet's mass has been reduced to about 1 $M_{\bigoplus}$ (panel b). The remaining mass is almost uncompressed; its mean density stays constant, so that the mass-radius exponent of the planet catches up with that of the Roche lobe, $\zeta_{\mathrm{p}} \simeq \zeta_{\mathrm{L}}$. With the decreasing planet's mass, at the stage $M_{\mathrm{p}} < 1.0 \ M_{\bigoplus}$, $\eta$ and mass transfer rate start to decrease (see panel b and d of Fig. \ref{mtsilicate}). The orbit does not change much any more (panel c). The evolution does not reach the stage where mass loss through L2 becomes significant. Further evolution would determined by tidal dissipation only. We stopped the calculation at $M_{\mathrm{p}} \simeq 0.5 M_{\bigoplus}$. 

The polytropic equation of state in Savonije's (1978) description of Roche lobe overflow is not a good approximation for `rocky' planets (even in their molten state close to the host star). The high surface density of such a planet should cause the onset of mass transfer to be much sharper than with the polytropic assumption used here. This will be a smaller effect in the later stages, in particular the transition to stable transfer will still be present. The dashed part of the lines in Fig.\ \ref{mtsilicate} is meant to indicate the uncertainty involved.

\section{Discussion and conclusions}
\label{summary}

We have investigated the mass transfer from a planet to its host star, with emphasis on the conditions under which Roche lobe overflow becomes dynamically unstable. The main factors are the adiabatic mass-radius relation of the planet and the processes redistributing angular momentum. Previous analyses of mass transfer were based on the analogy with mass-transferring binary systems (CVs and X-ray binaries) where the moment of inertia of the host star can be neglected, and the angular momentum of the transferred mass is returned to the orbit by tidal interaction with an accretion disc\footnote{ In the \textsc{mesa} code, the default value `Ritter' for the parameter `mdot\_scheme' assumes this, as in Ritter (1988).}. We have argued that this is not an appropriate assumption in the case of Roche lobe overflow of a planet on a MS  (or larger) star. The orbital angular momentum of a planet is far too small to affect the rotation of such stars, which can effectively act as an arbitrary sink of angular momentum. The consequence is that planets are much more likely to go through dynamically unstable Roche lobe overflow than predicted by the standard description that applies to interacting binaries.

The more massive planets touch the host star surface before overflowing their Roche lobes. These planets consequently spiral into their host on a short time-scale (we call this the `direct merger' case). This is especially likely to happen to planets (if any) that formed close to their host during their PMS life (see Figs.\ \ref{solargiant}-\ref{solargiants7} and \ref{tin1}, \ref{tin2}).

At some what lower planetary masses or larger radius of host stars, the mass-transferring stream intersects the surface of the star before completing an orbit. In this simple case (which we call the `stream impact' case), the specific angular momentum of the mass lost by the planet through the L1 is a bit less than that of the planet's centre of mass. This would cause a slight expansion of the orbit. In section \ref{discplanet} we have argued that even if the stream does not impact directly, the truncation radius of the prospective accretion disc is probably only marginally outside the stellar surface of a MS host star. Instead of forming a dics, we find it more likely that the hydrodynamic interaction with the slowly rotating stellar envelope is effective enough to absorb most the accreting angular momentum instead of leading to a spreading dics. We have called this the `minimal assumption', but have compared its consequences with different assumptions.

Whether Roche lobe overflow of a planet is dynamically stable and unstable depends on its (adiabatic) mass-radius relation, which differs strongly for gas giant planets and rocky planets (see Figs.\ \ref{zamsgiant}, \ref{zamssilicate}). The areas in the $M_{\rm p}-M_*$ plane differ correspondingly. For low mass rocky planets, the mass-radius index is close to the value $1/3$ for a nearly constant mean density. Since this is also the approximate mass-radius index of the Roche lobe of a low-mass companion, Roche lobe overflow is near marginal stability for the `minimal assumption' (see Fig.\ \ref{silicams}). The more massive rocks are slightly compressible. As a result, their mass transfer is initially unstable, but settles to stable overflow when the mass has decreased to less than $1\,M_{\bigoplus}$ (Fig.\ \ref{mtsilicate}). 

The higher likelihood of dynamically unstable Roche lobe overflow in our `minimal assumption' increases the change of observing a planet in the process of rapid mass transfer. The low transfer rates driven only by tidal interaction in stable overflow may be hard to recognize observationally.

\section*{Acknowledgements} 

We would like to thank Achim Weiss for the stellar models used in this work and the anonymous referee for very detailed and insightful comments that helped to improve the manuscript. SJ acknowledges support from the MPG-CAS Joint Doctoral Promotion Program (DPP) and Max Planck Institute for Astrophysics (MPA).

\bibliographystyle{mnras}

\label{lastpage}

\end{document}